\documentclass[crop]{CSL}

\usepackage{aas_macros}

\newcommand{\pdf}{\mathrm{Pr}}

\providecommand{\url}[1]{\href{#1}{#1}}
\providecommand{\dodoi}[1]{doi:~\href{http://doi.org/#1}{\nolinkurl{#1}}}
\providecommand{\doeprint}[1]{\href{http://ascl.net/#1}{\nolinkurl{http://ascl.net/#1}}}
\providecommand{\doarXiv}[1]{\href{https://arxiv.org/abs/#1}{\nolinkurl{https://arxiv.org/abs/#1}}}

\definecolor{quotecol}{rgb}{0.1,0.28,0.46}

\begin{document}

\supertitle{Research Paper}

\title[Do SETI Optimists Have a Fine-Tuning Problem?]{Do SETI Optimists Have a Fine-Tuning Problem?}

\author[Kipping \& Lewis]{David Kipping$^{1}$ and Geraint Lewis$^{2}$}

\address{\indent\add{1}{Department of Astronomy,
Columbia University,
550 W 120th Street,
New York, NY 10027, USA}\\
\indent\add{2}{Sydney Institute for Astronomy,
School of Physics, A28,
The University of Sydney,
NSW 2006, Australia}}

\corres{\name{David Kipping} \email{dkipping@astro.columbia.edu}}

\begin{abstract}
In ecological systems, be it a petri dish or a galaxy, populations evolve from some initial value (say zero) up to a steady state equilibrium, when the mean number of births and deaths per unit time are equal. 
This equilibrium point is a function of the birth and death rates, as well as the carrying capacity of the ecological system itself. We show that the occupation fraction versus birth-to-death rate ratio is S-shaped, saturating at the carrying capacity for large birth-to-death rate ratios and tending to zero at the other end. We argue that our astronomical observations appear inconsistent with a cosmos saturated with ETIs, and thus SETI optimists are left presuming that the true population is somewhere along the transitional part of this S-curve. Since the birth and death rates are a-priori unbounded, we argue that this presents a fine-tuning problem. Further, we show that if the birth-to-death rate ratio is assumed to have a log-uniform prior distribution, then the probability distribution of the ecological filling fraction is bi-modal - peaking at zero and unity. Indeed, the resulting distribution is formally the classic Haldane prior, conceived to describe the prior expectation of a Bernoulli experiment, such as a technological intelligence developing (or not) on a given world.
Our results formally connect the Drake Equation to the birth-death formalism, the treatment of ecological carrying capacity and their connection to the Haldane perspective.
\end{abstract}


\selfcitation{Kipping D \& Lewis G (2024).Do SETI Optimists Have a Fine-Tuning Problem? International Journal of Astrobiology (submitted). https://doi.org/10.1017/xxxxx}

\received{xx xxxx xxxx}

\revised{xx xxxx xxxx}

\accepted{xx xxxx xxxx}

\maketitle

\Fpagebreak

\section{The Haldane Perspective}
\label{sec:haldane1}

In 1968, Edwin \citeauthor{jaynes:1968} imagined that we are presented with a jar containing an unknown and unlabelled compound; call it chemical X. Consider that along a laboratory bench we find a large number of beakers filled with water and our task is to investigate how often chemical X will dissolve within them. Jaynes argued that one should reasonably expect the compound to either dissolve in nearly every instance, or almost never. It would be particularly remarkable if compound X were to dissolve approximately half of the time. Such an outcome would imply that the small variations in temperature, pressure, etc across the room were sufficient to tip the outcome either way. Whilst certainly possible, this scenario implies that the conditions in the room, and indeed the properties of the compound, were balanced on a knife edge; fine-tuned to yield such an outcome. Indeed, this argument trivially extends to cases where the conditions are extremely diverse, since again there's no reason why the range of conditions should saddle a switch-point that is a-priori unknown.

\citet{jaynes:1968} proposed that the a-priori probability distribution for the fraction of beakers ($F$) in which the compound will dissolve, lacking any other information, should $\propto F^{-1/2} (1-F)^{-1/2}$. Indeed, it can be shown that this is the Jeffrey's prior \citep{jeffreys:1946} of a Bernoulli process. Jaynes' prior has a bowl-like shape, peaking at the extreme values of $F=0$ and $F=1$, with a minimum at $F=0.5$. In fact, it was \citet{haldane:1923} who first introduced a prior distribution with this shape, but he instead suggested $\propto F^{-1} (1-F)^{-1}$. Haldane's prior is improper over the interval $[0,1]$, not normalizing to unity, but his proposal equally captures Jaynes' intuition regarding the outlined gedankenexperiment.

If we replace the beakers with planets, and the act of dissolving with the act of abiogenesis, it has been suggested that the Haldane prior is equally appropriate in an astrobiological context \citep{kipping:2020}. Consider an ensemble of Earth-like planets across the cosmos - worlds with similar gravity, composition, chemical inventories and climatic conditions. Although small differences will surely exist across space (like the beakers across the laboratory), one should reasonably expect that life either emerges nearly all of the time in such conditions, or hardly ever. As before, it would seem contrived for life to emerge in approximately half of the cases - again motivated from the fine-tuning perspective.

\begin{figure*}
\begin{center}
\includegraphics[width=16.0cm,angle=0,clip=true]{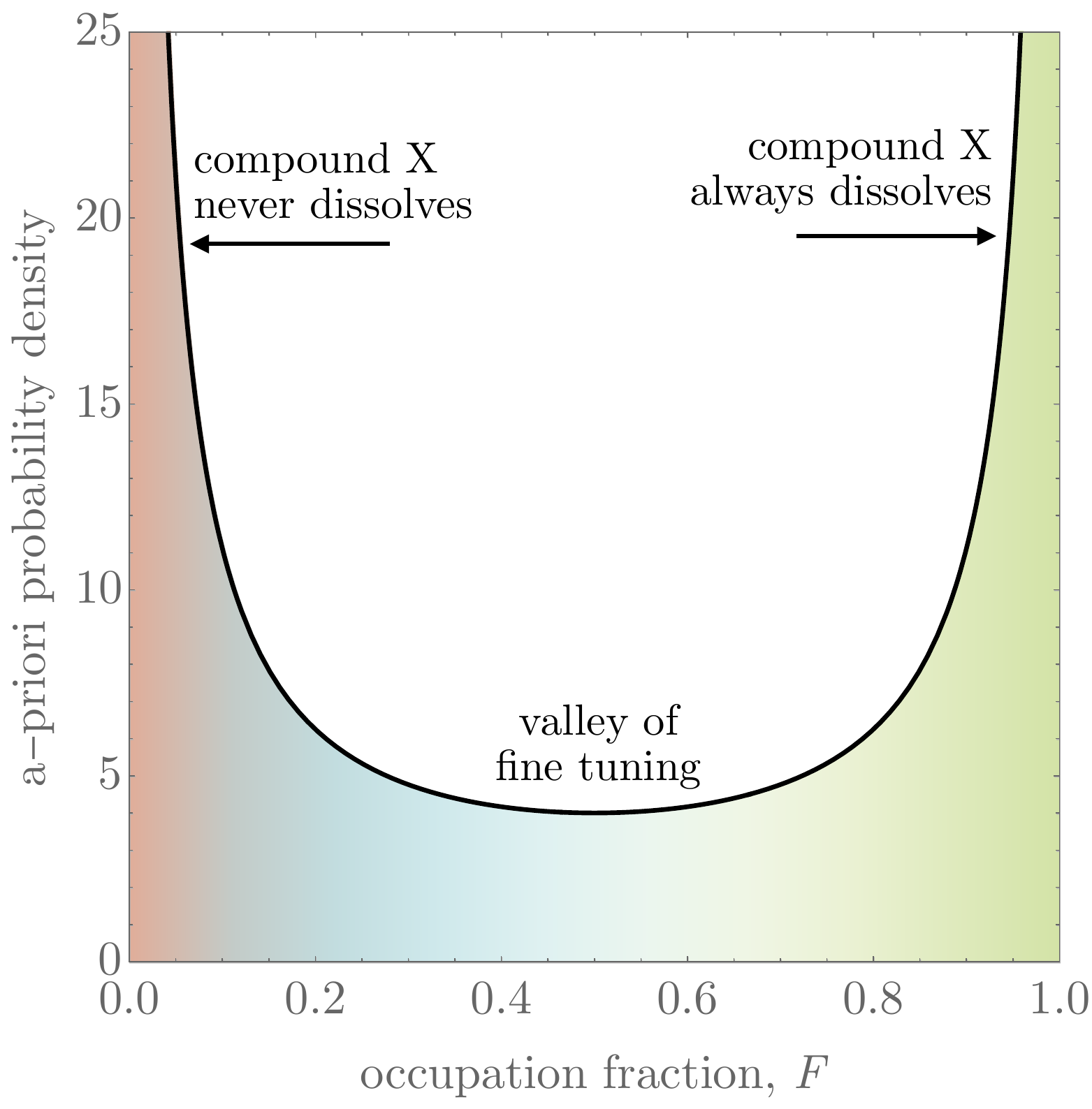}
\caption{\emph{
In the gedankenexperiment of attempting to dissolve an unknown compound X into a series of water vessels, Jaynes and Haldane argued that, a-priori, X will either dissolve almost all of the time or very rarely, but it would be contrived for nearly half of the cases to dissolve and half not. The function plotted here represents the Haldane prior ($F^{-1} (1-F)^{-1}$) that captures this behaviour.
}}
\label{fig:haldane}
\end{center}
\end{figure*}

This argument can be extended to other relevant astrobiology terms too, such as the fraction of worlds that are occupied with multicellular life, or technologically sophisticated species. As stated, the Haldane perspective seems persuasive, but perhaps a little qualitative and lacking rigorous justification. It's also unclear what it's implications really are to astrobiology, since all it really states is that the two extremes are equally likely - a lonely versus a crowded Universe.

One might be tempted to invoke our own existence as a data point here; indeed Jaynes wrote that a single experimental result can be used to produce a fairly conclusive determination of how the other experiments will fare. However, Jaynes was considering the case of an \textit{outside} observer, whose existential fate was divorced from the outcome as to whether compound X dissolved or not. This is manifestly not true here. We only exist because a success occurred. It is quite possible that successes are extremely rare, but we necessarily must be one of those successes as a self-aware sentient being. We are describing here an example of the weak anthropic principle \citep{carter:1974}, which demotes the observation that we exist to a completely useless datum - given that it's a conditional necessity.

Many authors have attempted to use the Copernican (or Mediocrity) Principle to argue for the plurality of life elsewhere in the cosmos (e.g. \citealt{rogers:2001,westby:2020}), but we caution against such a path. Consider for example the frequency of surface liquid water on a planet. Before observing any other planets within the Solar System (i.e. a-priori), one might posit that, by the Copernican Principle, liquid water must be typical on planetary surfaces. Such a claim is clearly wrong with the hindsight of modern observations. The reason that the Copernican Principle fails here is because our very existence is generally accepted to be dependent upon that water \citep{meadows:2018,schwieterman:2018,hallsworth:2021}. In contrast, consider the frequency of Neptune-like planets in the Universe. Neptune and Uranus do not appear to have any significant influence on Earth's development, and thus there's no obvious objection to invoking the Copernican Principle here. Indeed, such planets have been found to be very common \citep{bonfils:2013,dressing:2015,hsu:2019,bryson:2020}, demonstrating a successful and appropriate use of the principle.

How can we interpret Haldane's perspective to astrobiology then? What are its implications, and is there ultimately a more rigorous justification for its validity? To make progress, it is necessary to connect Haldane's perspective to Drake's - as in the Drake Equation \citep{drake:1965}. Regardless of the considerable and on-going debate concerning its utility (which we will touch on), the Drake Equation is the foundation upon which theoretical SETI\footnote{Search for Extraterrestrial Intelligence} rests. In this work, we will first bridge the Drake Equation to a more modern and concise formulation that dissolves the common criticisms levied at it. This also reveals how the problem of fine-tuning emerges for SETI optimists. We will then show how this modern formalism naturally leads to Haldane's perspective in and discuss the implications towards the end.

\section{The Steady State Drake Equation}
\label{sec:steady}

\subsection{The case for simplifying the Drake Equation}

The Drake Equation expresses the mean number of communicative civilizations in the Galaxy \citep{drake:1965}, given by
\begin{align}
N_C &= R_{\star} \times f_P \times n_E \times f_L \times f_I \times f_C \times L_C,
\end{align}
where $R_{\star}$	is	mean rate of star formation,
$f_P$ is the fraction of stars that have planets,
$n_E$ is the mean number of planets that could support life per star with planets,
$f_L$ is the fraction of life-supporting planets that develop life,
$f_I$ is the fraction of planets with life where life develops intelligence,
$f_C$ is the fraction of intelligent civilizations that develop communication and
$L_C$ is the mean length of time that civilizations can communicate.

We note that many terms frequently used in the SETI literature are problematic \citep{almar:2011}, including the term ``civilization'' for its Eurocentric implications \citep{denning:2011,decolonizing:2012}. In what follows, we favour the term ETIs, which although historically has been interpreted as ``Extra-Terrestrial Intelligences'', we here use to define ``Extraterrestrial Technological Instantiations''.

When discussing the Drake Equation, it's important to emphasise that $N_C$ is an expectation value. It does not represent the exact number of ETIs at any one time \citep{glade:2012}, but rather the average over some time interval over which the dependent parameters are approximately stable \citep{kipping:2021}.

A common pedagogical exercise is to ask students to guess the various values and estimate $N_C$ for themselves, and indeed these guesses are typically as reasonable as any other, since the latter four terms are wholly unknown \citep{sandberg:2018}. Indeed, this ignorance could be easily weaponized to dismiss the utility of the Drake Equation \citep{gertz:2021}, to which proponents often counter that it was never intended to be used as a calculator like this - the original intent was merely to convene a meeting and ``organize our ignorance'', to quote Jill Tarter \citep{achenbach:2000}.

Although once must concede that a full set of accurate inputs cannot be supplied to the Drake Equation \citep{sandberg:2018}, that does not mean it has no quantitative utility - the very framing of the problem implies certain statistical results. For example, \citet{maccone:2010} argued that the choice of which terms to include is somewhat debatable and one could reasonably conceive of longer lists of multiplicative parameters e.g. fraction of life forms that go on to develop multicellular life. If the list of parameters are treated as independently distributed then the Central Limit Theorem dictates that $N_C$ must follow a log-normal distribution, despite the fact we do not know what the probability distributions for each parameter even are.

\subsection{The birth-death formalism}

In our previous work \citep{kipping:2021}, it was suggested to imagine the reverse. Rather than trying to expand the Drake Equation ad infinitum \citep{maccone:2010}, one may compress it to the most efficient form possible. It can be seen that the first six parameters in the Drake Equation, including whatever additional intermediate terms one might wish to add (e.g. \citealt{molina:2019}), all describe the net process of spawning communicative ETIs - the birth rate\footnote{Note that the birth and death rates used here refer to average rates across the ensemble.}. Indeed, these six terms together have units of inverse time. The final term is conceptually different in describing the death process - how long the communicative ETI lasts. We may thus re-write the Drake Equation as simply
\begin{align}
N_C &= \underbrace{R_{\star} \times f_P \times n_E \times f_L \times f_I \times f_C}_{\equiv\Gamma_C} \times L_C.
\end{align}
This birth-death formalism is flexible and easily interpretable. For example, it's trivial to exchange the terms to cover different variations of $N$. For example, if $N_I$ is the number of ETIs (irrespective of their communicative intent), then $N_I = \Gamma_I L_I$. Further, it's also easy to change the volume from the entire Galaxy (\'a la Drake's original expression), to some relevant sample of interest (e.g. a specific cluster).

This formalism also dissolves many criticisms levied at the Drake Equation. First, there is no longer a problem with temporal retardation effects as pointed out by \citet{cirkovic:2004}. Whereas in the original framing it's unclear as to whether $R_{\star}$ should be the current star formation rate \citep{gertz:2021}, or the value from 4.4\,Gyr ago, or something in-between, that question is avoided completely by simply stating that there is \textit{some} current birth rate at which ETIs emerge, and that's that. Second, it resolves the terracentric tunnel vision effect one might critique the original equation with \citep{cirkovic:2007}, which considers a singular pathway to intelligence that is perhaps ignorant of other roads (e.g. tidally heated moons of rogue planets; \citealt{abbot:2011}). Indeed, the apparent subjectiveness of which terms to include and which to not (e.g. \citealt{molina:2019}) is moot here.

Finally, the formalism is indisputable. Worlds inhabited by species using technology must sometimes emerge (since we are one) and thus one can always define the number of newly birthed examples per unit time. The mechanism of this emergence is irrelevant in this framing, it could be via natural evolution \citep{kipping:2020}, directed panspermia \citep{crick:1973} or something else entirely - all pathways simply sum into this one birth rate. Similarly, one can always define the number of ETIs that terminate per unit time. Of course, these advantages come at the expense of a disadvantage - the birth-death formalism is not as useful for organizing a scientific conference, the original purpose of the Drake Equation \citep{gertz:2021}.

\subsection{Lessons from a previous attempt at the stationary Drake Equation}

In \citet{kipping:2021}, it was argued that the birth-death formalism of the Drake Equation allows one to say something about the distribution of the number of ETIs over time\footnote{In contrast, \citet{maccone:2010} consider the prior distribution of the mean number i.e. with respect to our subjective belief.}. Whereas $N_C$ represents the mean number of communicative ETIs, let's define $n_C$ as the time varying number - the stochastic value which should fluctuate around $N_C$. Accordingly, one may write that $\mathrm{E}[n_C]=N_C$. The birth process is characterized by a mean number of communicative ETIs birthed per unit time, $\Gamma_C$, which rigorously defines a Poisson process \citep{glade:2012}. For the death process, an exponential distribution was assumed in \citet{kipping:2021}, following the arguments made in \citet{contact:2020}. It was then demonstrated that the product of a Poisson and an exponential process yields another Poisson process. Hence, $n_C \sim \mathrm{Po}[N_C]$.

However, one feature of this approach seems in error and reveals an apparent weakness in the birth-death simplification - indeed a weakness that extends to the original Drake Equation too. The two terms, $\Gamma_C$ and $L_C$, have extreme permissible ranges \citep{lacki:2016} and thus could be varied such that $N_C$ exceeds the number of stars in the Galaxy (or whatever volume one is considering). This also rings true for the original Drake Equation, since $L_C$ is naively unbounded and could likewise be engineered to yield more ETIs than stars. This might seem like an edge effect; afterall, few would reasonably argue for a value of $L_C>10^{11}$\,years that would create this issue since this exceeds the age of the Universe \citep{planck:2018}. Nevertheless, it rigorously demonstrates that the Drake Equation is incomplete, missing some detail that leads to the correct asymptotic behaviour.

That something is \textit{carrying capacity}. In any environment, be is a petri dish or a galaxy, there is a finite carrying capacity of that volume to host the life forms under consideration - a fact understood in ecology since at least the 1840's \citep{sayre:2008}. Microbial colonies can't grow ad infinitum in a test tube. In what follows, we build upon our previous work \citep{kipping:2021} to account for this effect, within the birth-death paradigm.

\subsection{A Steady State Drake (SSD) Equation}

In this work, we concern ourselves with the steady state condition, where the mean number of births per unit time equals the mean number of deaths per unit time. This equilibrium point will of course drift over aeons due to the evolution of the cosmos (e.g. the star formation rate), but we assume that the extant population is in such a steady state. The conditions for this statement to hold are explicated later.

Let us write that the mean number of births per location over a time interval, $\delta t$, is given by $\lambda_B \delta t$. Note that already we have deviated from the parameterization of \citet{kipping:2021}, who consider the birth rate summed over all worlds ($\Gamma$), not \textit{per world} as done so here\footnote{Although both are per unit time.}. Also note our deliberately vaguely defined ``location'' phrasing above. To some degree, how one defines a location is up to the user and is inextricable with how one defines an occupied location. It could be planets, rocky objects, star systems, or even cubic parsecs. We'll use the term ``seats'' in what follows, as in potential seats for occupation.

Consider that there exists $N_A$ available seats, and thus we have an expectation value of $N_A \lambda_B \delta t$ births over the interval $\delta t$, where the actual birth number in any given time interval will vary stochastically about this expectation value following a Poisson distribution \citep{kipping:2021}. The number of available seats will not be equal to the total number of seats, $N_T$, since we assume here that a birth cannot occur within a seat that is already occupied. Hence, we write that $N_A=(N_T-N_O)$, where $N_O$ is the mean number of occupied seats.

\begin{figure*}
\begin{center}
\includegraphics[width=16.0cm,angle=0,clip=true]{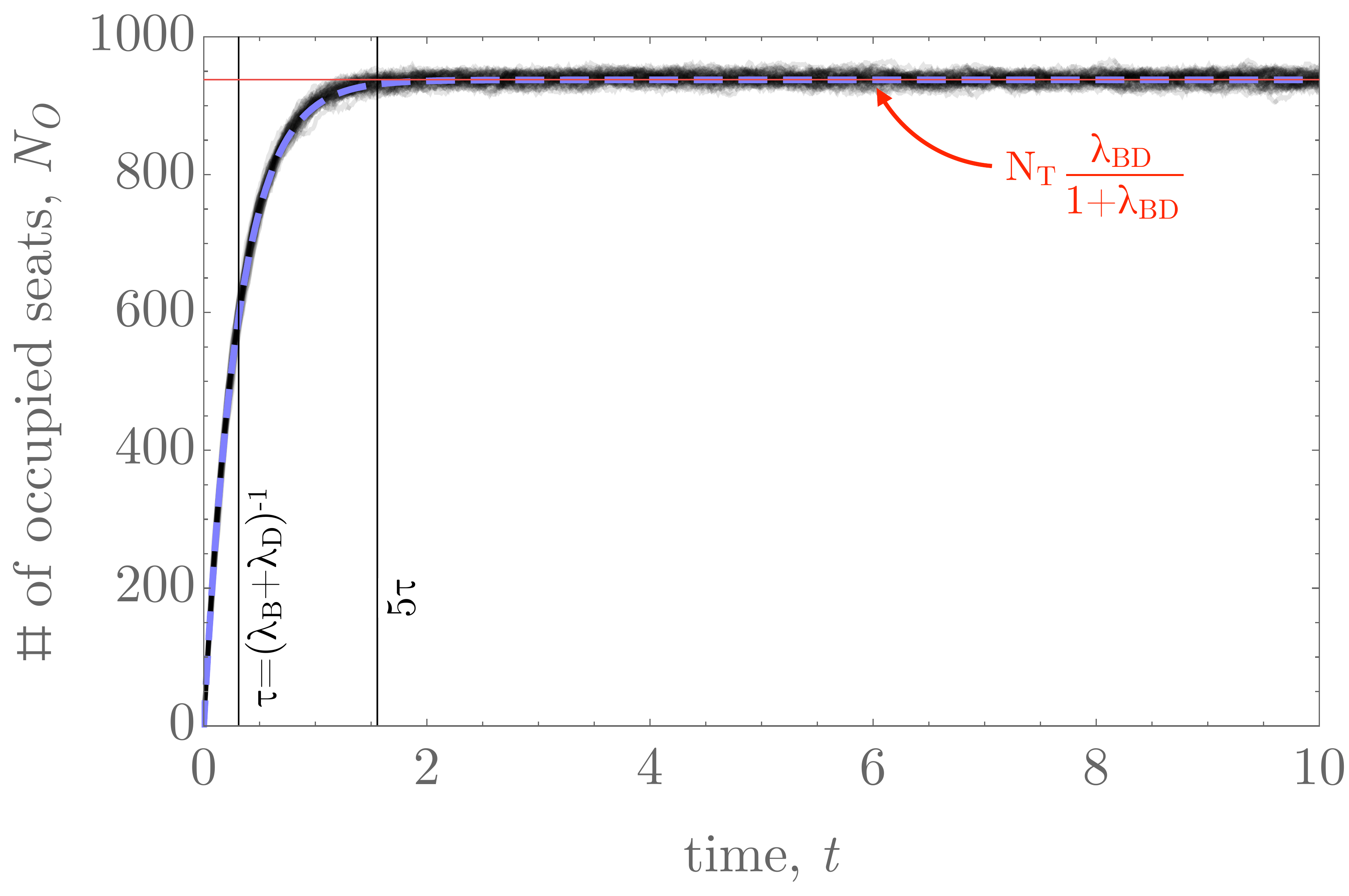}
\caption{\emph{
Using the parameters $\delta t=0.01$, $\lambda_B=3$, $\lambda_D=0.2$ for $N_T=1000$, we perform 50 Monte Carlo simulations of birth/death actions (wiggly black lines). The horizontal red line represents the steady state value of $N_O$, as predicted by Equation~(\ref{eqn:SSD}), which matches the simulations. The characteristic time folding time to reach steady state is given by $\tau$. Finally, the blue line shows the predictive growth towards steady state using Equation~(\ref{eqn:stabilization}).
}}
\label{fig:stabilization}
\end{center}
\end{figure*}

The mean number of deaths over a time interval, $\delta t$, will equal the mean number of occupied seats multiplied by the fraction that are expected to perish. Let us write that over a time interval $\delta t$, the fraction of occupied seats that will extinguish is $\lambda_D \delta t$. Thus $\lambda_D$ represents the death rate (of the current epoch)\footnote{Another way to arrive at this result is to assume a constant hazard function, such that lifetimes follow an exponential distribution and thus the fraction that die will be $1-e^{-\lambda_D \delta t}$. In the limit of small $\delta t$, this becomes $\lambda_D \delta t$. }. We may now balance the births and deaths as
\begin{align}
(N_T-N_O)\lambda_B &= N_O \lambda_D,
\end{align}
where the $\delta t$ time intervals have cancelled out. This expressions defines our condition for a steady state. Re-arranging, we obtain
\begin{align}
F &= \frac{ \lambda_{BD} }{ 1 + \lambda_{BD} },
\label{eqn:SSD}
\end{align}
where we have used the substitutions $F \equiv (N_O/N_T)$ (i.e. the occupation fraction) and $\lambda_{BD} \equiv (\lambda_B/\lambda_D)$. We refer to Equation~(\ref{eqn:SSD}) as the Steady State Drake (SSD) Equation in what follows. Figure~\ref{fig:stabilization} depicts 50 Monte Carlo simulations where we set $\delta t=0.01$, $\lambda_B=3$ and $\lambda_D=0.2$ for $N_T=1000$ total seats. Since we have a discrete number of seats and we know the mean number of births/deaths in each interval, then the stochastic realisations are simply drawn from binomial distributions. Using Equation~\ref{eqn:SSD}, we expect the simulation to stabilize at a steady state of $N_O=937.5$, marked by the horizontal red line.

Before continueing, we highlight some nuances in the above. First, we have only stated that there exists some rate at which ETIs are born, and correspondingly die. The mechanism of these processes is not relevant to the mere act of defining that a pair of such rates must exist. Noteably, it could happen via interaction or not; either way one can always still define such quantities. Second, our primary assumption is that a steady state condition exists. Again, this could occur via interaction models or not, our only assumption is that the population is in a steady state regardless. We will explore what it takes to violate this assumption later in the paper (such as certain interaction behaviours), as well as the consequences. Third, our model also assumes a finite number of seats for potential occupation, but this can readily be interpreted as the number of cubic parsecs, and thus can be defined in a way robust against manipulation. Put together, our reference to a ``carrying capacity'' refers to the finite number of seats, but we acknowledge that this can elicit confusion as limits to growth exist along other axes too, such as nutrition, resources and energy (which would all feed into the birth rate). Our only claim with respect to ``carrying capacity'' is that the number of occupied seats cannot exceed the number of available seats, a third assumption of this work.

\subsection{The optimist's fine-tuning problem}
\label{sub:finetuning}

\begin{figure*}
\begin{center}
\includegraphics[width=16.0cm,angle=0,clip=true]{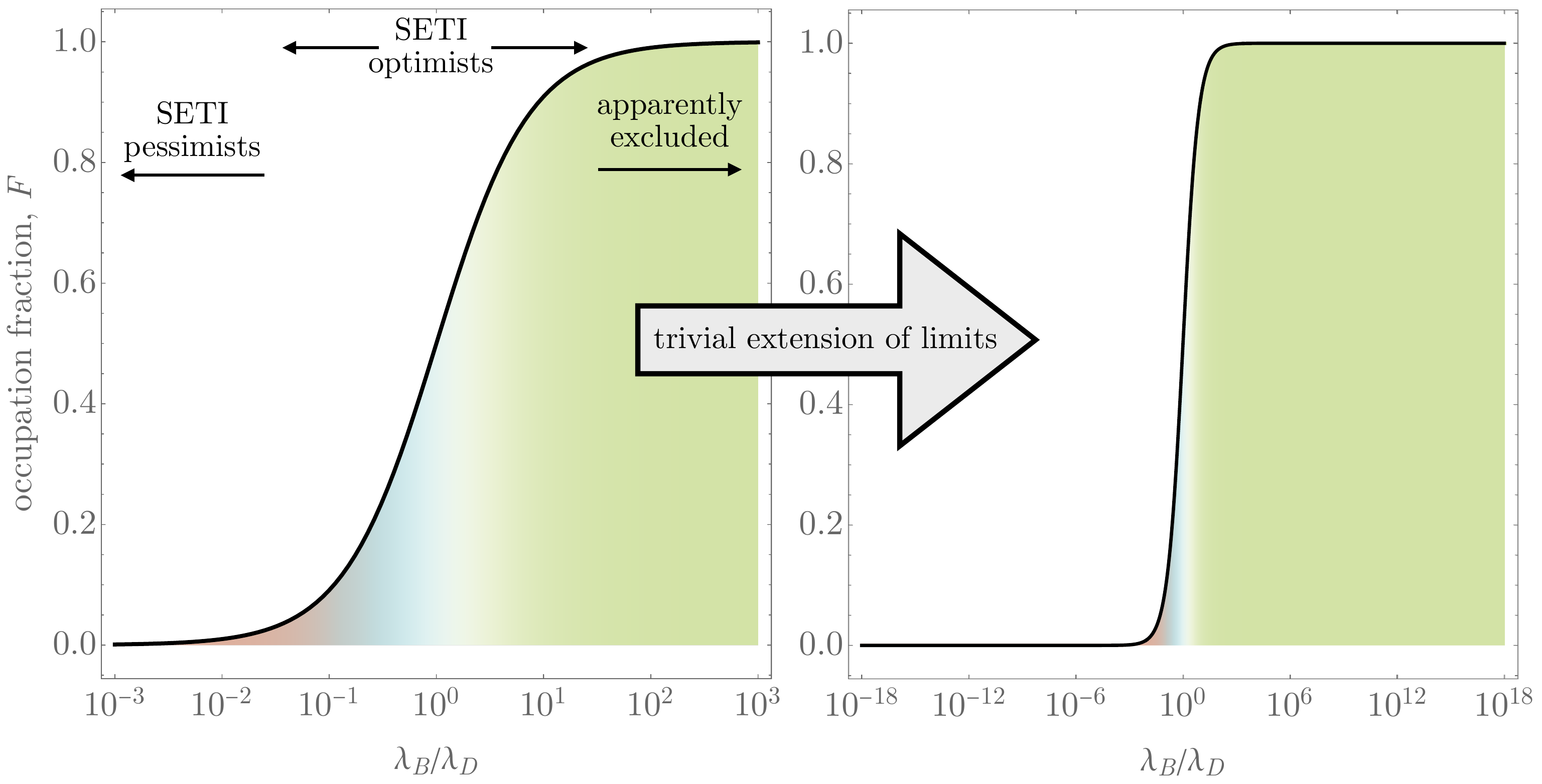}
\caption{\emph{
Left: Occupation fraction of potential ``seats'' as a function of the birth-to-death rate ratio ($\lambda_B/\lambda_D$), accounting for finite carrying capacity. In the context of communicative ETIs, an occupation fraction of $F\sim1$ is apparently incompatible with both Earth's history and our (limited) observations to date. Values of $\lambda_B/\lambda_D\ll1$ imply a lonely cosmos, and thus SETI optimists must reside somewhere along the middle of the $S$-shaped curve. Right: As we expand the bounds on $\lambda_B/\lambda_D$, the case for SETI optimism appears increasingly contrived and becomes a case of fine-tuning.
}}
\label{fig:S}
\end{center}
\end{figure*}

The term $\lambda_{BD}$ is unknown to us, and could in fact span many orders of magnitude \textit{a-priori} \citep{lacki:2016}. Plotting $F$ as a function of $\lambda_{BD}$ on a linear-log scale (see Figure~\ref{fig:S}), one sees a familiar $S$-shaped curve characteristic of population growth models with finite carrying capacity \citep{pearl:1920}. The saturation at high $\lambda_{BD}$ reflects the finite carrying capacity - increasing $\lambda_{BD}$ any further has little effect on $F$. In contrast, at low $\lambda_{BD}$, we asymptotically approach a lonely universe.

It is this curve that reveals the basic dilemma facing SETI optimists. We first establish that it would seem most improbable that $F\approx1$. Although our historical SETI surveys are woefully incomplete \citep{wright:2018}, there presently exists no reproducible, compelling evidence for other technological entities. Everything we observe about the cosmos appears consistent with this unsettling prospect. A counter argument is that we may be the subject of some grand conspiracy, namely the ``Zoo Hypothesis'' \citep{ball:1973}, which is potentially compatible with $F\approx1$. However, maintaining a monolithic culture at galactic scales given the finite speed of causality makes such a scenario highly contrived \citep{forgan:2017}.

More over, $F\approx1$ is simply incompatible with Earth's history. Most of Earth's history lacks even multicellular life, let alone a technological civilization\footnote{Although over timescales of Myr, lost civilizations are plausible; see \citet{wright:2018b} and \citet{schmidt:2019}.} (also see \citealt{kipping:2020}). We thus argue that $F\approx1$ can be reasonably dismissed as a viable hypothesis, which correspondingly excludes $\lambda_{BD} \gg 1$ by Equation~(\ref{eqn:SSD}) - although we revisit this point in Section~\ref{sub:F1}. 
We highlight that excluding $F\approx1$ is compatible with placing a ``Great Filter'' at any position, such as the ``Rare Earth'' hypothesis \citep{ward:2000} or some evolutionary ``Hard Step'' \citep{carter:2008}.

If one concedes that $F\not\approx1$, which we will take as granted in what follows, then what are the remaining scenarios? One lies somewhere along the steeply ascending $S$-curve, corresponding to $\lambda_{BD}\approx1$. This is what we consider to be the SETI optimist's scenario (given that $F\approx1$ is not allowed). Here, $F$ takes on modest but respectable values, sufficiently large that one might expect success with a SETI survey. For example, modern SETI surveys scan $N_T \sim 10^{3}$-$10^{4}$ targets (e.g. \citealt{enriquez:2017,maire:2019,price:2020,margot:2023,ma:2023}), so for such a survey to be successful one requires $F$ to exceed the reciprocal of this (i.e. $F \geq 10^{-4}$), but realistically greatly so (i.e. $F \gg 10^{-4}$) since not every occupied seat will produce the exact technosignature we are searching for, in the precise moments we look, and at the power level we are sensitive to. This arguably places the SETI optimist is a rather narrow corridor of requiring $N_T^{-1} \ll \lambda_{BD} \lesssim 1$.

The requirement for such fine-tuning forms the basis of our concern. It might be argued that as we increase $N_T$, the contrivance decreases and thus the fine-tuning problem dissolves. In a relative sense, we agree that the contrivance certainly diminishes as $N_T$ grows, but in an absolute sense the contrivance may still be extreme nonetheless. The basic problem is that $\lambda_{BD}$ has no clear lower limit, and could be plausibly be \textit{outrageously} small. For example, the probability of spontaneously forming proteins from amino acids has been estimated be ${\sim}10^{-77}$ \citep{axe:2004}, to say nothing of the many subsequent steps needed to produce living creatures - let alone technology development.

One might try to argue that our very existence demands $F \geq 1/N_{\star}$, where $N_{\star}$ is the number of stars in the observable Universe. But truly this limit is not justifiable, since the Universe appears to be much larger than the Hubble volume \citep{planck:2018}, and thus most Hubble volumes could be devoid of sentience - we necessarily live in the one where it occurred via the weak anthropic principle \citep{carter:1974}.

\section{Haldane's Return}
\label{sec:haldane2}

\subsection{Probabilistic steady state}

The SSD equation relates the occupation fraction, $F$, as a function of a single input parameter, $\lambda_{BD}$. However, it does not speak to the probability distribution of $F$. For that, we need to adopt some a-priori distribution for $\lambda_{BD}$ and then evaluate the implied distribution in $F$-space.

As a scale parameter, let's take the common approach of adopting a log-uniform prior on $\lambda_{BD}$, equivalent to a uniform prior on $\gamma_{BD} = \log\lambda_{BD}$. We bound $\gamma_{BD}$ between $\gamma_{BD,\mathrm{min}}$ and $\gamma_{BD,\mathrm{max}}$, such that
\begin{align}
\pdf(\gamma_{BD})\,\mathrm{d}\gamma_{BD} &= \frac{1}{\gamma_{BD,\mathrm{max}} - \gamma_{BD,\mathrm{min}}}\,\mathrm{d}\gamma_{BD}.
\label{eqn:gammaBD}
\end{align}
Since $F=\lambda_{BD}/(\lambda_{BD}+1)$ (Equation~\ref{eqn:SSD}), then by re-arrangement we have
\begin{align}
\lambda_{BD} &= \frac{F}{1-F},
\end{align}
and thus
%
\begin{align}
\gamma_{BD} &= \log\Big(\frac{F}{1-F}\Big).
\label{eqn:gamma2}
\end{align}
We can calculate the probability distribution of $F$ by noting that
\begin{align}
\pdf(F)\,\mathrm{d}F &= \pdf(\gamma_{BD})\,\mathrm{d}\gamma_{BD},\nonumber\\
\qquad&= \frac{1}{\gamma_{BD,\mathrm{max}} - \gamma_{BD,\mathrm{min}}}\,\frac{\mathrm{d}\gamma_{BD}}{\mathrm{d}F}\,\mathrm{d}F,
\label{eqn:Ftemp}
\end{align}
where we have used Equation~(\ref{eqn:gammaBD}) on line two. The derivative on the right-hand side is easily calculated from Equation~(\ref{eqn:gamma2}), giving
\begin{align}
\frac{\mathrm{d}\gamma_{BD}}{\mathrm{d}F} &= \frac{\mathrm{d}}{\mathrm{d}F} \log\Big(\frac{F}{1-F}\Big),\nonumber\\
\qquad&= \frac{1}{F(1-F)}.
\end{align}
Substituting this result into Equation~(\ref{eqn:Ftemp}), we finally obtain
\begin{align}
\pdf(F)\,\mathrm{d}F &= \frac{1}{\log\lambda_{BD,\mathrm{max}} - \log\lambda_{BD,\mathrm{min}}} \frac{1}{F (1-F)}\,\mathrm{d}F,
\label{eqn:PF}
\end{align}
which is essentially the Haldane prior. A subtle difference between Haldane's prior is that the original version lacks the normalization terms present here, and is defined over the interval $F$ of $[0,1]$, whereas here the bounds are $F_{\mathrm{min}}$ and $F_{\mathrm{max}}$, which can be found by Equation~(\ref{eqn:SSD}) to yield
\begin{align}
\pdf(F)\,\mathrm{d}F &= \frac{1}{ \log\big(\frac{F_{\mathrm{max}}}{1-F_{\mathrm{max}}}\big) - \log\big(\frac{F_{\mathrm{min}}}{1-F_{\mathrm{min}}}\big) } \frac{1}{F (1-F)}\,\mathrm{d}F.
\label{eqn:PF2}
\end{align}
Note that integrating the above from $F_{\mathrm{min}}$ to $F_{\mathrm{max}}$ equals unity. In other words our version of the Haldane prior is \textit{proper}, resolving the common criticism of the original Haldane prior.

\subsection{Haldane = SSD}

This derivation resolves the underlying connection between the Drake Equation and the Haldane perspective. Starting from the Haldane perspective, which seems intuitionally appropriate for the case of astrobiology as argued in Section~\ref{sec:haldane1}, we would conclude that the cosmos is either very lonely or very crowded, whereas as intermediate values appear contrived. Since a crowded cosmos appears incompatible with observations, SETI optimists find themselves in the fine-tuning valley (see Figure~\ref{fig:haldane}). This matches the conclusion found using the SSD equation - given the enormous potential range of the birth-to-death rate ratio, $\lambda_{BD}$, the cosmos is likely nearly empty or fully occupied, with intermediate states requiring fine-tuning. We have shown why these two appear to show the same thing, they are in fact equivalent under the assumption that $\lambda_B/\lambda_D$ is a-priori log-uniformly distributed - the generic uninformative prior of a scale parameter. In the Appendix, we show that one can also engineer a Jaynes prior to manifest, instead of the Haldane, by adopting a uniform prior in $-2\sin^{-1}(1+\lambda_{BD})^{-1/2}$ (instead of $\log\lambda_{BD}$). However, we argue there is no clear justification for favouring this over a log-uniform prior.

\subsection{Conditions for a steady state}
\label{sub:conditions}

To arrive at this result, we have assumed that the current population of extant ETIs exists in a steady state. We here investigate under what conditions this assumption is appropriate. If we initialize at some time $t=0$, the number of occupied seats will clearly not be in such a state and will require some time to grow and reach equilibrium. The relationship between the evolution of the occupied states is given by
\begin{align}
\frac{d N_O}{dt} = -\frac{d N_A}{dt},\nonumber\\
\qquad&= \lambda_B N_A - \lambda_D N_O,\nonumber\\
\qquad&= \lambda_B N_T - ( \lambda_B + \lambda_D ) N_O,
\end{align}
subject to the initial condition $N_O(0) = 0$. Solving this, it can be shown that this growth follows
\begin{align}
N_O(t) &= \Bigg(\frac{N_T \lambda_{BD}}{1 + \lambda_{BD}}\Bigg) \Bigg(1 - \exp\big( -(\lambda_B+\lambda_D)t\big) \Bigg).
\label{eqn:stabilization}
\end{align}
Accordingly, the timescale for settling into the steady state is several $\tau\equiv(\lambda_B+\lambda_D)^{-1}$. Recall that $\lambda_D^{-1}$ corresponds to the mean lifetime of occupation, and one might reasonably argue that this must be much smaller than the age of the Universe, $H_0^{-1}$, hence $\lambda_D^{-1} \ll H_0^{-1}$. If $\lambda_D^{-1} \ll \lambda_B^{-1}$, then this argument guarantees that the steady state condition has been satisfied already.

In reality, $\lambda_B$ and $\lambda_D$ will also vary over cosmic history due to the changing environment of the Universe itself, such metallicity enrichment and evolving star formation rates. However, if the characteristic timescale over which this cosmic evolution occurs is much greater than the equilibrium timescale ($\tau$), then the number of occupied seats will remain in a steady state, albeit one for which the equilibrium point slowly drifts.

\section{Salvaging Hope}
\label{sec:discussion}

Although our conclusion casts doubt on the chances of a successful SETI program, we argue that SETI is an important and vital experiment that deserves dedicated resources. Whilst the odds of success appear small, such a success would arguably represent the most impactful scientific discovery in human history \citep{wright:2023}. Further, there are several ways of salvaging hope in our formalism that we describe in what follows.

\subsection{Paths to optimism within the SSD framework?}

First, we currently have no lower bound on $\lambda_{BD}$ and thus the fact it can be arbitrarily (even outrageously) small underpins the entire fine-tuning argument made here. As shown earlier in Section~\ref{sub:finetuning},
the SETI optimist requires that $N_T^{-1}\ll\lambda_{BD}\lesssim1$, but current knowledge is fully compatible with $\lambda_{BD,\mathrm{min}} \lll N_T^{-1}$ \citep{lacki:2016} - hence the conclusion that SETI optimists live in a narrow corridor. Accordingly, one path to salvage hope would be to place a firm lower limit on $\lambda_{BD}$ that is comparable to, or greater than $N_T^{-1}$ i.e. we require $\lambda_{BD,\mathrm{min}} \gtrsim N_T^{-1}$. In practice, it's difficult to imagine how we could ever place a constraint on this parameter besides a successful SETI detection -  even null detections have little power here since there are innumerable ways in which a ETI could be missed \citep{kipping:2024}.

Of course, we are always free to increase $N_T$ i.e. perform ever larger SETI surveys. This will increase the relative odds of success in direct proportion to $N_T$. However, without a lower bound on $\lambda_{BD,\mathrm{min}}$ this doesn't, in general, lead to optimism in an absolute sense. For example, consider that $\lambda_{BD} = 10^{-30}$. In this case, a case perfectly compatible with everything we know about our cosmos, increasing $N_T$ from say a value of $10^4$ typical of modern SETI surveys (e.g. \citealt{enriquez:2017,maire:2019,price:2020,margot:2023,ma:2023}) to every star in the galaxy ($10^{11}$) increases the relative probability of success by 10 million. However, despite this increase, the absolute probability of success is still minuscule, just $10^{-19}$. In contrast, the suppose $\lambda_{BD} = 10^{-10}$ instead, which again is equally plausibly, then the expectation value becomes $10$ technospheres amongst the ensemble. Of course, these are just examples and a rigorous quantification would require knowledge of shapes (e.g. \citet{jeffreys:1946}, \citet{lacki:2016}, etc) and bounding limits (e.g. see \citealt{spiegel:2012}) of a prior on $\lambda_{BD}$, over which one could then marginalize to compute expectation values. A core argument of this work is that there is no sensible lower limit to $\lambda_{BD}$ with present information, and thus it extends down into the abyss of extremely and arbitrarily small values.

The SSD equation only involves these two parameters, $\lambda_{BD}$ and $N_T$, yet neither provides the leverage to dismiss the fine-tuning argument in a practical sense. Under the assumptions of this work, efforts to salvage optimism must thus turn to violating either 1) the SSD itself 2) the assumption that $F\not\approx1$. We believe these are only the only two pathways to retaining plausible optimism for SETI.

\subsection{Violating the SSD}

Consider 1) first - violating the SSD. As explored earlier in Section~\ref{sub:conditions},
the steady state condition is not guaranteed and strictly assumes that the elapsed time within the window of cosmic habitability exceeds several $\tau\equiv(\lambda_B+\lambda_D)^{-1}$. If this is false, equilibrium has not yet been achieved and the number of occupied seats is presumably in ascent. A decline is also possible, but this assumes the conditions of the cosmos are more hostile to life now than previously, which is difficult to justify. An ascent would be compatible with some roaming ETI colonizing the galaxy \citep{bracewell:1960,freitas:1980,tipler:1980}, thus themselves engineering $\lambda_B$ exponentially upwards over time. However, we note that the settlement front is expected to fill the Galaxy $\lesssim$\,Gyr even for ``slow'' probes (30\,km/s), aided by the motion of the stars themselves \citep{carroll:2019}. Accordingly, there is arguably another fine-tuning problem that we find ourselves living during during the mid-point of this relatively rapid transition - too early and there's no-one to talk to, too late and we shouldn't be here to think about it. A similar argument can be made for any process in which $\lambda_B$ undergoes rapid growth (e.g. directed panspermia; \citealt{crick:1973}). If, instead, slow growth is favoured, then we fall back into the steady state condition.

\subsection{Allowing for $F\approx1$}
\label{sub:F1}

An alternative solution is 2) - violating the assumption that $F\not\approx1$. Such a position is challenged by our observations of the local Universe and why it was treated as forbidden in much of this work. However, a possible way around this is to invoke the ``Grabby Aliens'' hypothesis \citep{hanson:2021} and the weak anthropic principle \citep{carter:1974}. Here, one might imagine that ETIs emerge rarely, but when they do, they often proceed to colonize their surrounding region in short order. In such a Universe, most regions are filled and thus $F\approx1$. The fact that we don't see $F\approx1$ locally is because humanity must necessarily have emerged in a volume of space where this wave has not yet reached, via the weak anthropic principle. We note that this solution falls into tension with the Cosmological Principle \citep{keel:2007}. Such a scenario lends itself to inverting the normal view of SETI - rather than looking locally, we should be looking at regions greatly separated from us. Such a hypothesis has the advantage that it is, in principle, verifiable via extragalactic SETI \citep{zackrisson:2015,griffith:2015}.

There are other ways to allow for $F\approx1$, such as invoking a universal technological ceiling. This is not a ``Great Filter'', since that speaks to the death rate. Instead, we have here a universal limit to technological development for all ETIs, and that limit is coincidentally not far off humanity's current level of advancement. In this picture, the fact that we see no evidence of technosignatures littering the sky is because ETIs never develop to a point where their footprint would be noticeable. In our view, this is a strong, contrived and unfounded assumption to assert. Similarly, co-ordinated ETI behaviour to avoid contact (the ``Zoo Hypothesis''; \citealt{ball:1973}) is challenged by the finite speed of light, the sheer scale of the Galaxy and likely heterogeneity of emergent behaviours \citep{forgan:2017}. Similar challenges exist for ``Dark Forest'' arguments \citep{yu:2015}, with the added issue as to why Earth was sterilized at some point in its long 4.5\,Gyr history already.

In summary, the case of $F\approx1$ may be in tension with our observations of the local Universe when considering SETI, but it is perfectly consistent with our observations when considering life more generally, especially if we treat the seats as star systems or Earth-like planets. By the SSD Equation, we would here argue for precisely the Haldane perspective - a cosmos either teeming with life or almost devoid of it, both of which are compatible with current observations. We therefore emphasize that our argument for pessimism does not extend to life more broadly.

\textit{Conflicts of Interest}: None

\ack[Acknowledgement]{
DK thanks the following for their generous support to the Cool Worlds Lab:
Douglas Daughaday,
Elena West,
Tristan Zajonc,
Alex de Vaal,
Mark Elliott,
Stephen Lee,
Zachary Danielson,
Chad Souter,
Marcus Gillette,
Tina Jeffcoat,
Jason Rockett,
Tom Donkin,
Andrew Schoen,
Reza Ramezankhani,
Steven Marks,
Nicholas Gebben,
Mike Hedlund,
Ryan Provost,
Nicholas De Haan,
Emerson Garland,
Queen Rd Fndn Inc
Scott Thayer,
Frank Blood,
Ieuan Williams,
Xinyu Yao \&
Axel Nimmerjahn
}




\onecolumn
\section*{Appendix}

We have shown in Section~\ref{sec:haldane2} that if one takes the SSD equation (Equation~\ref{eqn:SSD}) and assumes that $\lambda_{BD}$ follows a log-uniform prior distribution, then $F$ follows the Haldane prior ($\propto F^{-1} (1-F)^{-1}$). As noted in Section~\ref{sec:haldane1}, the sibling to the Haldane prior is the Jaynes prior ($\propto F^{-1/2} (1-F)^{-1/2}$), which exhibits similar qualitative behaviour in $F$-space but is a proper prior over the interval $[0,1]$. In addition, the Jaynes prior is rigorously the Jeffrey's prior for a Bernoulli process \citep{jeffreys:1946}. Accordingly, one might wonder, how does $\lambda_{BD}$ need to be distributed to cause $F$ to follow the Jaynes prior, instead of the Haldane?

By reverse engineering the derivation shown in Section~\ref{sec:haldane2}, we find that one needs to assume $-2\sin^{-1}(1+\lambda_{BD})^{-1/2}=\zeta$ is uniformly distributed to cause this behaviour - since this is the anti-derivative of $F^{-1/2}(1-F)^{-1/2}$. To see this, consider first that $\zeta$ is uniformly distributed such that

\begin{align}
\pdf(\zeta)\,\mathrm{d}\zeta &= \frac{1}{\zeta_{\mathrm{max}}-\zeta_{\mathrm{min}}}\,\mathrm{d}\zeta.
\end{align}

By continuity of probability, we thus have

\begin{align}
\pdf(F)\,\mathrm{d}F &= \frac{1}{\zeta_{\mathrm{max}}-\zeta_{\mathrm{min}}}\frac{\mathrm{d}\zeta}{\mathrm{d}F}\,\mathrm{d}F,\nonumber\\
\qquad&= \frac{1}{\zeta_{\mathrm{max}}-\zeta_{\mathrm{min}}} \frac{1}{\sqrt{F}\sqrt{1-F}} \,\mathrm{d}F.
\end{align}

The limits can now be replaced by virtue of our $\zeta$ definition to give

\begin{align}
\pdf(F)\,\mathrm{d}F &= \frac{1}{2\sin^{-1}(\sqrt{1-F_{\mathrm{min}}}) - 2\sin^{-1}(\sqrt{1-F_{\mathrm{max}}})}
\frac{1}{\sqrt{F}\sqrt{1-F}} \,\mathrm{d}F,
\end{align}

which integrates to unity over the interval $[F_{\mathrm{min}},F_{\mathrm{max}}]$, as expected.

Having established this produces the desired behaviour, it is instuctive to convert this prior into $\lambda_{BD}$ space natively by noting that

\begin{align}
\pdf(\lambda_{BD})\,\mathrm{d}\lambda_{BD} &= \frac{1}{\zeta_{\mathrm{max}}-\zeta_{\mathrm{min}}} \frac{\mathrm{d}\zeta}{\mathrm{d}\lambda_{BD}} \,\mathrm{d}\lambda_{BD}.
\end{align}

It is easy to show that

\begin{align}
\frac{\mathrm{d}\zeta}{\mathrm{d}\lambda_{BD}} &= \frac{\mathrm{d}}{\mathrm{d}\lambda_{BD}} \Bigg( -2\sin^{-1}(1+\lambda_{BD})^{-1/2} \Bigg),\nonumber\\
\qquad&= \frac{1}{\sqrt{\lambda_{BD}}(1+\lambda_{BD})},
\end{align}

and thus

\begin{align}
\pdf(\lambda_{BD})\,\mathrm{d}\lambda_{BD} &= \frac{1}{\zeta_{\mathrm{max}}-\zeta_{\mathrm{min}}} \frac{1}{\sqrt{\lambda_{BD}}(1+\lambda_{BD})} \,\mathrm{d}\lambda_{BD},\nonumber\\
\qquad&= \frac{1}{
2\sin^{-1}(1+\lambda_{\mathrm{min}})^{-1/2} - 2\sin^{-1}(1+\lambda_{\mathrm{max}})^{-1/2}
} \frac{1}{\sqrt{\lambda_{BD}}(1+\lambda_{BD})} \,\mathrm{d}\lambda_{BD}.
\label{eqn:jayneslin}
\end{align}

Following a similar derivation, one can show that

\begin{align}
\pdf(\log\lambda_{BD})\,\mathrm{d}\log\lambda_{BD} &\propto \frac{\sqrt{e^{\log\lambda_{BD}}})}{(1+e^{\log\lambda_{BD}})} \,\mathrm{d}\log\lambda_{BD}.
\label{eqn:jayneslog}
\end{align}

\begin{figure}
\begin{center}
\includegraphics[width=16.4cm,angle=0,clip=true]{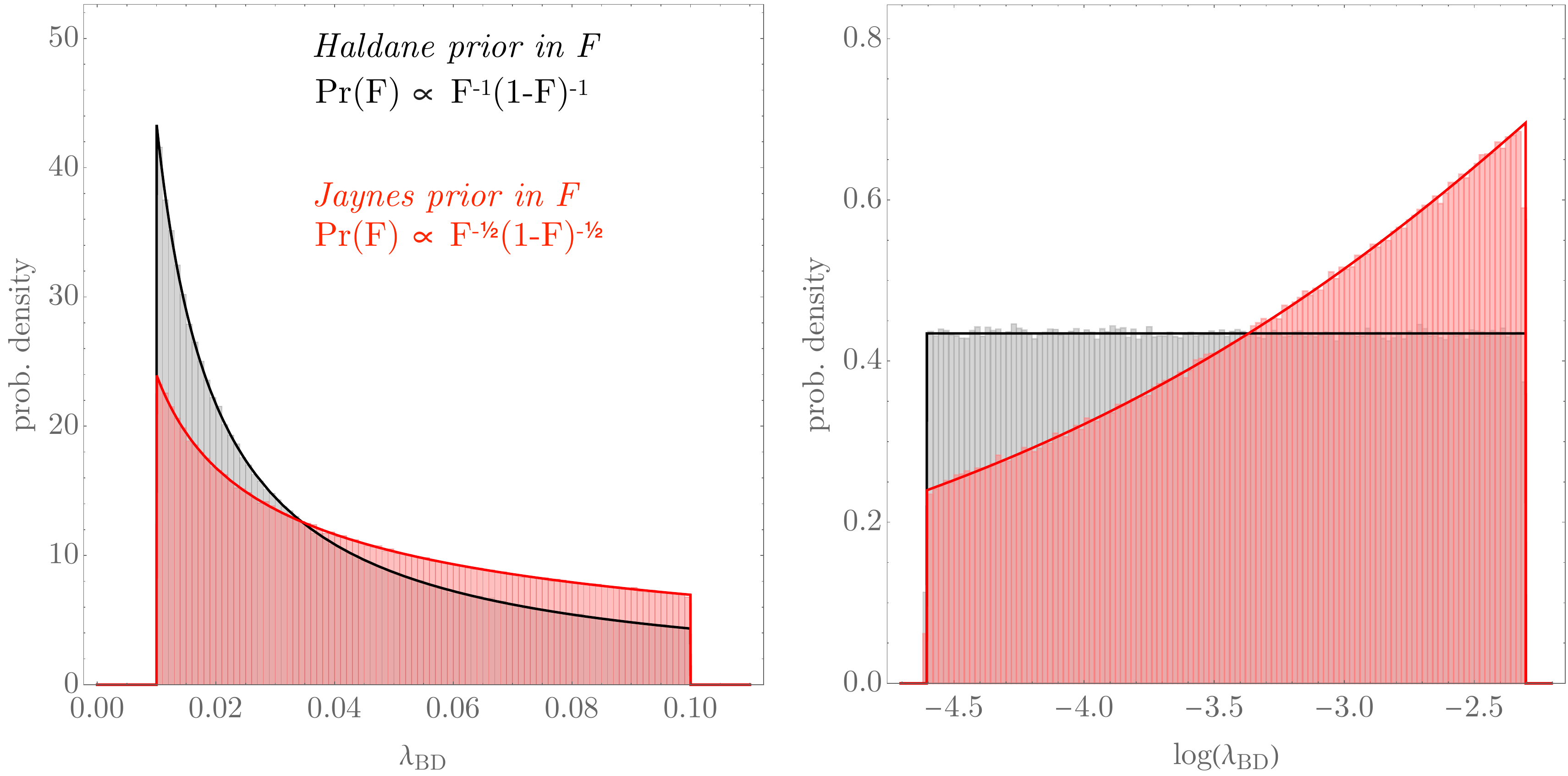}
\caption{\emph{
Left: Probability distribution of $\lambda_{BD}$ assuming $F=\lambda_{BD}/(1+\lambda_{BD})$ (Equation~\ref{eqn:SSD}) follows a Haldane (black) versus Jaynes (red) prior. The histograms show the result of $10^6$ Monte Carlo samples, where we assume $\lambda_{BD,\mathrm{min}}=0.01$ and $\lambda_{BD,\mathrm{max}}=0.1$. Right: Same as left but in $\log\lambda_{BD}$ space.
}}
\label{fig:jaynes}
\end{center}
\end{figure}

We plot Equation~(\ref{eqn:jayneslin}) \& (\ref{eqn:jayneslog}) in the left and right panels of Figure~\ref{fig:jaynes}, respecitvely, in red. Alongside, we plot the same functions but assuming the Haldane prior in $F$ in black, which is to say a log-uniform prior in $\lambda_{BD}$. This reveals thier similar but distinct behaviour. In particular, the Jaynes prior places greater weight at large $\lambda_{BD}$ values. Since $\lambda_{BD}$ is a scale parameter with some unknown minimum and maximum, a log-uniform prior is the most uninformative choice thus justifying our decision to adopt it in this paper.

\begin{thebibliography}{}
\bibitem[{{Abbot} \& {Switzer}(2011)}]{abbot:2011}
{Abbot}, D.~S., \& {Switzer}, E.~R. 2011, \apjl, 735, L27,
  \dodoi{10.1088/2041-8205/735/2/L27}

\bibitem[{{Achenbach}(2020)}]{achenbach:2000}
{Achenbach}, J. 2020, National Geographic, 197, 24

\bibitem[{Almár(2011)}]{almar:2011}
Almár, I. 2011, Acta Astronautica, 68, 351,
  \dodoi{https://doi.org/10.1016/j.actaastro.2009.07.010}

\bibitem[{Axe(2004)}]{axe:2004}
Axe, D.~D. 2004, Journal of Molecular Biology, 341, 1295,
  \dodoi{https://doi.org/10.1016/j.jmb.2004.06.058}

\bibitem[{{Ball}(1973)}]{ball:1973}
{Ball}, J.~A. 1973, \icarus, 19, 347, \dodoi{10.1016/0019-1035(73)90111-5}

\bibitem[{{Bonfils} {et~al.}(2013){Bonfils}, {Delfosse}, {Udry}, {Forveille},
  {Mayor}, {Perrier}, {Bouchy}, {Gillon}, {Lovis}, {Pepe}, {Queloz}, {Santos},
  {S{\'e}gransan}, \& {Bertaux}}]{bonfils:2013}
{Bonfils}, X., {Delfosse}, X., {Udry}, S., {et~al.} 2013, \aap, 549, A109,
  \dodoi{10.1051/0004-6361/201014704}

\bibitem[{{Bracewell}(1960)}]{bracewell:1960}
{Bracewell}, R.~N. 1960, \nat, 186, 670, \dodoi{10.1038/186670a0}

\bibitem[{Bryson {et~al.}(2020)Bryson, Coughlin, Batalha, Berger, Huber, Burke,
  Dotson, \& Mullally}]{bryson:2020}
Bryson, S., Coughlin, J., Batalha, N.~M., {et~al.} 2020, The Astronomical
  Journal, 159, 279, \dodoi{10.3847/1538-3881/ab8a30}

\bibitem[{{Carroll-Nellenback} {et~al.}(2019){Carroll-Nellenback}, {Frank},
  {Wright}, \& {Scharf}}]{carroll:2019}
{Carroll-Nellenback}, J., {Frank}, A., {Wright}, J., \& {Scharf}, C. 2019, \aj,
  158, 117, \dodoi{10.3847/1538-3881/ab31a3}

\bibitem[{{Carter}(1974)}]{carter:1974}
{Carter}, B. 1974, in Confrontation of Cosmological Theories with Observational
  Data, ed. M.~S. {Longair}, Vol.~63, 291--298

\bibitem[{{Carter}(2008)}]{carter:2008}
{Carter}, B. 2008, International Journal of Astrobiology, 7, 177,
  \dodoi{10.1017/S1473550408004023}

\bibitem[{{{\'C}irkovi{\'c}}(2004)}]{cirkovic:2004}
{{\'C}irkovi{\'c}}, M.~M. 2004, Astrobiology, 4, 225,
  \dodoi{10.1089/153110704323175160}

\bibitem[{{{\'C}irkovi{\'c}}(2007)}]{cirkovic:2007}
---. 2007, International Journal of Astrobiology, 6, 325,
  \dodoi{10.1017/S1473550407003916}

\bibitem[{{Crick} \& {Orgel}(1973)}]{crick:1973}
{Crick}, F.~H.~C., \& {Orgel}, L.~E. 1973, \icarus, 19, 341,
  \dodoi{10.1016/0019-1035(73)90110-3}

\bibitem[{Denning(2011)}]{denning:2011}
Denning, K. 2011, Acta Astronautica, 68, 381,
  \dodoi{https://doi.org/10.1016/j.actaastro.2009.11.019}

\bibitem[{{Drake}(1965)}]{drake:1965}
{Drake}, F.~D. 1965, in In Current aspects of exobiology (Elsevier Science),
  323--345

\bibitem[{{Dressing} \& {Charbonneau}(2015)}]{dressing:2015}
{Dressing}, C.~D., \& {Charbonneau}, D. 2015, \apj, 807, 45,
  \dodoi{10.1088/0004-637X/807/1/45}

\bibitem[{{Enriquez} {et~al.}(2017){Enriquez}, {Siemion}, {Foster}, {Gajjar},
  {Hellbourg}, {Hickish}, {Isaacson}, {Price}, {Croft}, {DeBoer}, {Lebofsky},
  {MacMahon}, \& {Werthimer}}]{enriquez:2017}
{Enriquez}, J.~E., {Siemion}, A., {Foster}, G., {et~al.} 2017, \apj, 849, 104,
  \dodoi{10.3847/1538-4357/aa8d1b}

\bibitem[{{Forgan}(2017)}]{forgan:2017}
{Forgan}, D.~H. 2017, International Journal of Astrobiology, 16, 349,
  \dodoi{10.1017/S1473550416000392}

\bibitem[{{Freitas}(1980)}]{freitas:1980}
{Freitas}, R.~A., J. 1980, Journal of the British Interplanetary Society, 33,
  251

\bibitem[{{Gertz}(2021)}]{gertz:2021}
{Gertz}, J. 2021, arXiv e-prints, arXiv:2105.03984,
  \dodoi{10.48550/arXiv.2105.03984}

\bibitem[{{Glade} {et~al.}(2012){Glade}, {Ballet}, \& {Bastien}}]{glade:2012}
{Glade}, N., {Ballet}, P., \& {Bastien}, O. 2012, International Journal of
  Astrobiology, 11, 103, \dodoi{10.1017/S1473550411000413}

\bibitem[{{Griffith} {et~al.}(2015){Griffith}, {Wright}, {Maldonado}, {Povich},
  {Sigur{\dj}sson}, \& {Mullan}}]{griffith:2015}
{Griffith}, R.~L., {Wright}, J.~T., {Maldonado}, J., {et~al.} 2015, \apjs, 217,
  25, \dodoi{10.1088/0067-0049/217/2/25}

\bibitem[{{Haldane}(1932)}]{haldane:1923}
{Haldane}, J.~B.~S. 1932, Proceedings of the Cambridge Philosophical Society,
  28, 55, \dodoi{10.1017/S0305004100010495}

\bibitem[{{Hallsworth} {et~al.}(2021){Hallsworth}, {Koop}, {Dallas}, {Zorzano},
  {Burkhardt}, {Golyshina}, {Mart{\'\i}n-Torres}, {Dymond}, {Ball}, \&
  {McKay}}]{hallsworth:2021}
{Hallsworth}, J.~E., {Koop}, T., {Dallas}, T.~D., {et~al.} 2021, Nature
  Astronomy, 5, 665, \dodoi{10.1038/s41550-021-01391-3}

\bibitem[{{Hanson} {et~al.}(2021){Hanson}, {Martin}, {McCarter}, \&
  {Paulson}}]{hanson:2021}
{Hanson}, R., {Martin}, D., {McCarter}, C., \& {Paulson}, J. 2021, \apj, 922,
  182, \dodoi{10.3847/1538-4357/ac2369}

\bibitem[{{Hsu} {et~al.}(2019){Hsu}, {Ford}, {Ragozzine}, \&
  {Ashby}}]{hsu:2019}
{Hsu}, D.~C., {Ford}, E.~B., {Ragozzine}, D., \& {Ashby}, K. 2019, \aj, 158,
  109, \dodoi{10.3847/1538-3881/ab31ab}

\bibitem[{Jaynes(1968)}]{jaynes:1968}
Jaynes, E.~T. 1968, IEEE Transactions on Systems Science and Cybernetics, 4,
  227, \dodoi{10.1109/TSSC.1968.300117}

\bibitem[{{Jeffreys}(1946)}]{jeffreys:1946}
{Jeffreys}, H. 1946, Proceedings of the Royal Society of London Series A, 186,
  453, \dodoi{10.1098/rspa.1946.0056}

\bibitem[{{Keel}(2007)}]{keel:2007}
{Keel}, W.~C. 2007, {The road to galaxy formation} ({Springer Praxis Books})

\bibitem[{{Kipping}(2020)}]{kipping:2020}
{Kipping}, D. 2020, Proceedings of the National Academy of Science, 117, 11995,
  \dodoi{10.1073/pnas.1921655117}

\bibitem[{{Kipping}(2021)}]{kipping:2021}
---. 2021, Research Notes of the American Astronomical Society, 5, 44,
  \dodoi{10.3847/2515-5172/abeb7b}

\bibitem[{{Kipping} {et~al.}(2020){Kipping}, {Frank}, \&
  {Scharf}}]{contact:2020}
{Kipping}, D., {Frank}, A., \& {Scharf}, C. 2020, International Journal of
  Astrobiology, 19, 430, \dodoi{10.1017/S1473550420000208}

\bibitem[{{Kipping} \& {Wright}(2024)}]{kipping:2024}
{Kipping}, D., \& {Wright}, J. 2024, \aj, 167, 24,
  \dodoi{10.3847/1538-3881/ad0cbe}

\bibitem[{{Lacki}(2016)}]{lacki:2016}
{Lacki}, B.~C. 2016, arXiv e-prints, arXiv:1609.05931,
  \dodoi{10.48550/arXiv.1609.05931}

\bibitem[{{Ma} {et~al.}(2023){Ma}, {Ng}, {Rizk}, {Croft}, {Siemion}, {Brzycki},
  {Czech}, {Drew}, {Gajjar}, {Hoang}, {Isaacson}, {Lebofsky}, {MacMahon}, {de
  Pater}, {Price}, {Sheikh}, \& {Worden}}]{ma:2023}
{Ma}, P.~X., {Ng}, C., {Rizk}, L., {et~al.} 2023, Nature Astronomy, 7, 492,
  \dodoi{10.1038/s41550-022-01872-z}

\bibitem[{{Maccone}(2010)}]{maccone:2010}
{Maccone}, C. 2010, Acta Astronautica, 67, 1366,
  \dodoi{10.1016/j.actaastro.2010.05.003}

\bibitem[{{Maire} {et~al.}(2019){Maire}, {Wright}, {Barrett}, {Dexter},
  {Dorval}, {Duenas}, {Drake}, {Hultgren}, {Isaacson}, {Marcy}, {Meyer},
  {Ramos}, {Shirman}, {Siemion}, {Stone}, {Tallis}, {Tellis}, {Treffers}, \&
  {Werthimer}}]{maire:2019}
{Maire}, J., {Wright}, S.~A., {Barrett}, C.~T., {et~al.} 2019, \aj, 158, 203,
  \dodoi{10.3847/1538-3881/ab44d3}

\bibitem[{{Margot} {et~al.}(2023){Margot}, {Li}, {Pinchuk}, {Myhrvold},
  {Lesyna}, {Alcantara}, {Andrakin}, {Arunseangroj}, {Baclet}, {Belk},
  {Bhadha}, {Brandis}, {Carey}, {Cassar}, {Chava}, {Chen}, {Chen}, {Cheng},
  {Cimbri}, {Cloutier}, {Combitsis}, {Couvrette}, {Coy}, {Davis}, {Delcayre},
  {Du}, {Feil}, {Fu}, {Gilmore}, {Grahill-Bland}, {Iglesias}, {Juneau},
  {Karapetian}, {Karfakis}, {Lambert}, {Lazbin}, {Li}, {Li}, {Liskij},
  {Lopilato}, {Lu}, {Ma}, {Mathur}, {Minasyan}, {Muller}, {Nasielski},
  {Nguyen}, {Nicholson}, {Niemoeller}, {Ohri}, {Padhye}, {Penmetcha},
  {Prakash}, {Qi}, {Rindt}, {Sahu}, {Scally}, {Scott}, {Seddon}, {Shohet},
  {Sinha}, {Sinigiani}, {Song}, {Stice}, {Tabucol}, {Uplisashvili}, {Vanga},
  {Vazquez}, {Vetushko}, {Villa}, {Vincent}, {Waasdorp}, {Wagaman}, {Wang},
  {Wight}, {Wong}, {Yamaguchi}, {Zhang}, {Zhao}, \& {Lynch}}]{margot:2023}
{Margot}, J.-L., {Li}, M.~G., {Pinchuk}, P., {et~al.} 2023, \aj, 166, 206,
  \dodoi{10.3847/1538-3881/acfda4}

\bibitem[{{Meadows} \& {Barnes}(2018)}]{meadows:2018}
{Meadows}, V.~S., \& {Barnes}, R.~K. 2018, in Handbook of Exoplanets, ed. H.~J.
  {Deeg} \& J.~A. {Belmonte} (Springer), 57,
  \dodoi{10.1007/978-3-319-55333-7_57}

\bibitem[{{Molina Molina}(2019)}]{molina:2019}
{Molina Molina}, J.~A. 2019, arXiv e-prints, arXiv:1912.01783,
  \dodoi{10.48550/arXiv.1912.01783}

\bibitem[{Pearl \& Reed(1920)}]{pearl:1920}
Pearl, R., \& Reed, L.~J. 1920, Proceedings of the National Academy of
  Sciences, 6, 275, \dodoi{10.1073/pnas.6.6.275}

\bibitem[{{Planck Collaboration} {et~al.}(2020){Planck Collaboration},
  {Aghanim}, {Akrami}, {Ashdown}, {Aumont}, {Baccigalupi}, {Ballardini},
  {Banday}, {Barreiro}, {Bartolo}, {Basak}, {Battye}, {Benabed}, {Bernard},
  {Bersanelli}, {Bielewicz}, {Bock}, {Bond}, {Borrill}, {Bouchet}, {Boulanger},
  {Bucher}, {Burigana}, {Butler}, {Calabrese}, {Cardoso}, {Carron},
  {Challinor}, {Chiang}, {Chluba}, {Colombo}, {Combet}, {Contreras}, {Crill},
  {Cuttaia}, {de Bernardis}, {de Zotti}, {Delabrouille}, {Delouis}, {Di
  Valentino}, {Diego}, {Dor{\'e}}, {Douspis}, {Ducout}, {Dupac}, {Dusini},
  {Efstathiou}, {Elsner}, {En{\ss}lin}, {Eriksen}, {Fantaye}, {Farhang},
  {Fergusson}, {Fernandez-Cobos}, {Finelli}, {Forastieri}, {Frailis},
  {Fraisse}, {Franceschi}, {Frolov}, {Galeotta}, {Galli}, {Ganga},
  {G{\'e}nova-Santos}, {Gerbino}, {Ghosh}, {Gonz{\'a}lez-Nuevo}, {G{\'o}rski},
  {Gratton}, {Gruppuso}, {Gudmundsson}, {Hamann}, {Handley}, {Hansen},
  {Herranz}, {Hildebrandt}, {Hivon}, {Huang}, {Jaffe}, {Jones}, {Karakci},
  {Keih{\"a}nen}, {Keskitalo}, {Kiiveri}, {Kim}, {Kisner}, {Knox},
  {Krachmalnicoff}, {Kunz}, {Kurki-Suonio}, {Lagache}, {Lamarre}, {Lasenby},
  {Lattanzi}, {Lawrence}, {Le Jeune}, {Lemos}, {Lesgourgues}, {Levrier},
  {Lewis}, {Liguori}, {Lilje}, {Lilley}, {Lindholm}, {L{\'o}pez-Caniego},
  {Lubin}, {Ma}, {Mac{\'\i}as-P{\'e}rez}, {Maggio}, {Maino}, {Mandolesi},
  {Mangilli}, {Marcos-Caballero}, {Maris}, {Martin}, {Martinelli},
  {Mart{\'\i}nez-Gonz{\'a}lez}, {Matarrese}, {Mauri}, {McEwen}, {Meinhold},
  {Melchiorri}, {Mennella}, {Migliaccio}, {Millea}, {Mitra},
  {Miville-Desch{\^e}nes}, {Molinari}, {Montier}, {Morgante}, {Moss}, {Natoli},
  {N{\o}rgaard-Nielsen}, {Pagano}, {Paoletti}, {Partridge}, {Patanchon},
  {Peiris}, {Perrotta}, {Pettorino}, {Piacentini}, {Polastri}, {Polenta},
  {Puget}, {Rachen}, {Reinecke}, {Remazeilles}, {Renzi}, {Rocha}, {Rosset},
  {Roudier}, {Rubi{\~n}o-Mart{\'\i}n}, {Ruiz-Granados}, {Salvati}, {Sandri},
  {Savelainen}, {Scott}, {Shellard}, {Sirignano}, {Sirri}, {Spencer},
  {Sunyaev}, {Suur-Uski}, {Tauber}, {Tavagnacco}, {Tenti}, {Toffolatti},
  {Tomasi}, {Trombetti}, {Valenziano}, {Valiviita}, {Van Tent}, {Vibert},
  {Vielva}, {Villa}, {Vittorio}, {Wandelt}, {Wehus}, {White}, {White},
  {Zacchei}, \& {Zonca}}]{planck:2018}
{Planck Collaboration}, {Aghanim}, N., {Akrami}, Y., {et~al.} 2020, \aap, 641,
  A6, \dodoi{10.1051/0004-6361/201833910}

\bibitem[{{Price} {et~al.}(2020){Price}, {Enriquez}, {Brzycki}, {Croft},
  {Czech}, {DeBoer}, {DeMarines}, {Foster}, {Gajjar}, {Gizani}, {Hellbourg},
  {Isaacson}, {Lacki}, {Lebofsky}, {MacMahon}, {Pater}, {Siemion}, {Werthimer},
  {Green}, {Kaczmarek}, {Maddalena}, {Mader}, {Drew}, \& {Worden}}]{price:2020}
{Price}, D.~C., {Enriquez}, J.~E., {Brzycki}, B., {et~al.} 2020, \aj, 159, 86,
  \dodoi{10.3847/1538-3881/ab65f1}

\bibitem[{{Rogers}(2001)}]{rogers:2001}
{Rogers}, N.~S. 2001, Journal of the British Interplanetary Society, 54, 424

\bibitem[{{Sandberg} {et~al.}(2018){Sandberg}, {Drexler}, \&
  {Ord}}]{sandberg:2018}
{Sandberg}, A., {Drexler}, E., \& {Ord}, T. 2018, arXiv e-prints,
  arXiv:1806.02404, \dodoi{10.48550/arXiv.1806.02404}

\bibitem[{Sayre(2008)}]{sayre:2008}
Sayre, N.~F. 2008, Annals of the Association of American Geographers, 98, 120,
  \dodoi{10.1080/00045600701734356}

\bibitem[{{Schmidt} \& {Frank}(2019)}]{schmidt:2019}
{Schmidt}, G.~A., \& {Frank}, A. 2019, International Journal of Astrobiology,
  18, 142, \dodoi{10.1017/S1473550418000095}

\bibitem[{{Schwieterman} {et~al.}(2018){Schwieterman}, {Kiang}, {Parenteau},
  {Harman}, {DasSarma}, {Fisher}, {Arney}, {Hartnett}, {Reinhard}, {Olson},
  {Meadows}, {Cockell}, {Walker}, {Grenfell}, {Hegde}, {Rugheimer}, {Hu}, \&
  {Lyons}}]{schwieterman:2018}
{Schwieterman}, E.~W., {Kiang}, N.~Y., {Parenteau}, M.~N., {et~al.} 2018,
  Astrobiology, 18, 663, \dodoi{10.1089/ast.2017.1729}

\bibitem[{Smith(2012)}]{decolonizing:2012}
Smith, L. 2012, Decolonizing Methodologies: Research and Indigenous Peoples
  (Otago University Press).
\newblock \url{https://books.google.com/books?id=41rYMgEACAAJ}

\bibitem[{Spiegel \& Turner(2012)}]{spiegel:2012}
Spiegel, D.~S. \& Turner, E.~L. 2012, PNAS, 109, 395, \dodoi{10.1073/pnas.1111694108}

\bibitem[{{Tipler}(1980)}]{tipler:1980}
{Tipler}, F.~J. 1980, \qjras, 21, 267

\bibitem[{{Ward} \& {Brownlee}(2000)}]{ward:2000}
{Ward}, P., \& {Brownlee}, D. 2000, {Rare earth : why complex life is uncommon
  in the universe} (Copernicus)

\bibitem[{{Westby} \& {Conselice}(2020)}]{westby:2020}
{Westby}, T., \& {Conselice}, C.~J. 2020, \apj, 896, 58,
  \dodoi{10.3847/1538-4357/ab8225}

\bibitem[{{Wright}(2018)}]{wright:2018b}
{Wright}, J.~T. 2018, International Journal of Astrobiology, 17, 96,
  \dodoi{10.1017/S1473550417000143}

\bibitem[{Wright {et~al.}(2023)Wright, Haramia, \& Swiney}]{wright:2023}
Wright, J.~T., Haramia, C., \& Swiney, G. 2023, Space Policy, 63, 101517,
  \dodoi{https://doi.org/10.1016/j.spacepol.2022.101517}

\bibitem[{{Wright} {et~al.}(2018){Wright}, {Kanodia}, \& {Lubar}}]{wright:2018}
{Wright}, J.~T., {Kanodia}, S., \& {Lubar}, E. 2018, \aj, 156, 260,
  \dodoi{10.3847/1538-3881/aae099}

\bibitem[{{Yu} {et~al.}(2015){Yu}}]{yu:2015}
{Yu}, C. 2015, JBIS, 68, 142

\bibitem[{{Zackrisson} {et~al.}(2015){Zackrisson}, {Calissendorff}, {Asadi}, \&
  {Nyholm}}]{zackrisson:2015}
{Zackrisson}, E., {Calissendorff}, P., {Asadi}, S., \& {Nyholm}, A. 2015, \apj,
  810, 23, \dodoi{10.1088/0004-637X/810/1/23}
\end{thebibliography}
\end{document}